\documentclass[11pt,epsf]{article}
\usepackage{comment}
\newtheorem{theorem}{Theorem}

\newcommand {\hbx} {\mbox{\boldmath $\hat{x}$}}

\newcommand {\hx} {\hat{x}}
\newcommand {\hX} {\hat{X}}

\newcommand {\dfn} {\stackrel{\Delta} {=}}
\newcommand {\exe} {\stackrel{\cdot} {=}}

\newcommand {\lexe} {\stackrel{\cdot} {\le}}

\newcommand{\lea}{\stackrel{\mbox{\tiny (a)}}{\le}}
\newcommand{\leb}{\stackrel{\mbox{\tiny (b)}}{\le}}

\newcommand {\reals} {{\rm I\!R}}

\newcommand {\bx} {\mbox{\boldmath $x$}}

\newcommand {\bE} {\mbox{\boldmath $E$}}

\newcommand {\bX} {\mbox{\boldmath $X$}}

\newcommand{\calB}{{\cal B}}
\newcommand{\calC}{{\cal C}}

\newcommand{\calE}{{\cal E}}

\newcommand{\calI}{{\cal I}}

\newcommand{\calP}{{\cal P}}

\newcommand{\calS}{{\cal S}}
\newcommand{\calT}{{\cal T}}

\newcommand{\calX}{{\cal X}}

\begin{document}
\thispagestyle{empty}
\title{A Universal Random Coding Ensemble\\
for Sample-wise Lossy Compression}
\author{Neri Merhav}
\date{}
\maketitle

\begin{center}
The Andrew \& Erna Viterbi Faculty of Electrical Engineering\\
Technion - Israel Institute of Technology \\
Technion City, Haifa 32000, ISRAEL \\
E--mail: {\tt merhav@ee.technion.ac.il}\\
\end{center}
\vspace{1.5\baselineskip}
\setlength{\baselineskip}{1.5\baselineskip}

\begin{abstract}
We propose a universal ensemble for random selection of rate-distortion codes,
which is asymptotically optimal in a sample-wise sense. According to this
ensemble, each reproduction vector, $\hbx$, is selected independently at random under the
probability distribution that is proportional to $2^{-LZ(\hbx)}$, where
$LZ(\hbx)$ is
the code-length of $\hbx$ pertaining to the 1978 version of the Lempel-Ziv (LZ)
algorithm. We show that, with high probability, the resulting codebook gives
rise to an asymptotically optimal
variable-rate lossy compression scheme under an arbitrary distortion measure,
in the sense that
a matching converse theorem also holds. According to the converse theorem,
even if the decoder knew
$\ell$-th order type of source vector in advance ($\ell$ being a large but
fixed positive integer), the performance of the
above-mentioned code could not have been improved essentially, for the vast
majority of codewords that represent all source vectors in the same
type. Finally, we provide a discussion of our results, which includes, among
other things, a comparison to a coding scheme that selects the
reproduction vector with the shortest LZ 
code length among all vectors that are within the allowed distortion from the source
vector.\\

\end{abstract}

\section{Introduction}

We revisit the well-known problem of lossy source coding for finite-alphabet
sequences with respect to (w.r.t.) a certain distortion
measure \cite{Berger71},
\cite[Chap.\ 10]{CT06}, \cite[Chap.\
9]{Gallager68}, \cite{Gray90}, \cite[Chaps.\ 7,8]{VO79}. More concretely,
our focus is on $d$-semifaithful codes, namely, variable--rate codes that meet
a certain distortion constraint for every source sequence (and not only in expectation).
As is very well known \cite{Berger71}, the rate-distortion function
quantifies the minimum achievable expected coding rate for a given
memoryless source and distortion measure.

During several past decades, many research efforts were motivated by the fact that the
source statistics are rarely (if not never)
known in practice, and were therefore
directed to the quest for
universal coding schemes, namely, coding schemes which do not depend of the
unknown statistics,
but nevertheless, approach the lower bounds (i.e., the entropy, in 
lossless compression, or the rate-distortion function, in the lossy case)
asymptotically, as the block length grows without bound.
We next provide a very brief (and non-comprehensive) review of
some of the relevant earlier works. 

In lossless compression, the theory of universal source coding
is very well developed and mature. Davisson's work \cite{Davisson73} concerning
universal-coding redundancies
has established the concepts of weak universality and strong 
universality (vanishing maximin and minimax redundancies, respectively), 
and has characterized the connection to the capacity of the
`channel' defined by family of conditional distributions of the data to be
compressed given the index (or parameter) of the source
in the class
\cite{Gallager76}. For many of the frequently encountered parametric
classes of sources,
the minimum achievable redundancy of universal codes is well-known to be
dominated by $\frac{k\log n}{2n}$, where $k$ is the
number of degrees of freedom of the parameter, and
$n$ is the block length. A central idea that arises from Davisson's
theory is to construct a Shannon code pertaining to
the probability distribution of the data 
vector w.r.t.\
a mixture (with a certain prior function) of all sources in
the class.
Rissanen, which was the inventor of the minimum description length (MDL) principle, has
proved in \cite{Rissanen84} a converse to a coding
theorem, which asserts that asymptotically, no universal code can achieve
redundancy below $(1-\epsilon)\frac{k\log n}{2n}$, with the possible
exception of sources from
a subset of the parameter space, whose volume tends to zero as $n\to\infty$,
for every positive $\epsilon$.
Merhav and Feder \cite{MF95} have generalized this result to more general
classes of sources, with the term $\frac{k\log n}{2n}$
substituted by the capacity of the above mentioned `channel'. 
Further developments, including more refined redundancy analyses, have been
carried out in later studies.

In the wider realm of universal lossy compression, the theory is,
unfortunately, not as sharp and well-developed as in the lossless setting.
We confine our attention, in this work, to
$d$-semifaithful codes \cite{OS90}, namely,
codes that
satisfy the distortion requirement with probability one.
Zhang, Yang and Wei \cite{ZYW97} have proved that, unlike in 
lossless compression, in the lossy case, even if the source
statistics are known perfectly, it is impossible to achieve
redundancy below $\frac{\log n}{2n}$ (see also \cite{me93}),
but $\frac{\log n}{n}$ is achievable. Not knowing the source conveys the price of
enlarging the multiplicative constant in front of $\frac{\log
n}{n}$. Indeed,
Yu and Speed \cite{YS93} have established weak universality with a constant that
grows with the cardinalities of the alphabets of the source and the reconstruction
\cite{SP21}. Ornstein and Shields \cite{OS90} have considered universal
$d$-semifaithful coding for stationary and ergodic sources w.r.t.\ the Hamming
distortion measure, and established convergence with probability one to the
rate-distortion function. Kontoyiannis \cite{Kontoyiannis00}
had several interesting findings. 
The first is a certain central limit theorem (CLT), with a $O(1/\sqrt{n})$
redundancy term, whose
coefficient is described as a limiting Gaussian random variable
with some constant variance. The second is the so called law of iterated
logarithm (LIL)
with redundancy proportional to $\sqrt{\frac{\log(\log n)}{n}}$
infinitely often with probability one. One of the counter-intuitive
conclusions from
\cite{Kontoyiannis00} is that universality is priceless under these
performance measures. In \cite{KZ02}, many of
the findings are based on the observation that optimal compression
can be characterized in terms of the negative logarithm of the probability of
a sphere of radius $nD$ around the source vector w.r.t.\ the distortion
measure, where $D$ is the allowed per-letter distortion.
In the same article, they proposed also the ensemble of random coding w.r.t.\
a probability distribution given by a mixture of all
distributions in a certain class.
In two recent articles, Mahmood and Wagner
\cite{MW22a}, \cite{MW22b} have studied $d$-semifaithful codes that are strongly
universal w.r.t.\ both the source and the distortion function.
The redundancy rates in \cite{MW22a} behave like $\frac{\log n}{n}$
with different
multiplicative constants.

A parallel line of research work on universal lossless and lossy compression, which was pioneered by Ziv, pertains to the
individual-sequence approach. According to this approach, there are no
assumptions on the statistical properties of the source. The source sequence
to be compressed is considered an arbitrary
deterministic (individual) sequence, but limitations are imposed on the
encoder and/or the decoder to be implementable by finite--state machines. This
includes, first and foremost, the celebrated Lempel-Ziv (LZ) algorithm
\cite{Ziv78}, \cite{ZL78}, 
as well as further developments that extend the scope to lossy compression
with and without side information \cite{MZ06}, \cite{Ziv84}, as well as to joint
source--channel coding \cite{me14}, \cite{me21}, \cite{me22}, \cite{Ziv80}.
In the lossless case, the article \cite{WMF94} provides
an individual-sequence analogue of the above-mentioned result due to Rissanen,
where the expression $\frac{k\log n}{2n}$
continues to designate the best achievable redundancy, but the main term of
the compression ratio there is the
empirical entropy of the source vector instead of the ordinary entropy of the
probabilistic setting. The
converse bound of \cite{WMF94} applies to the vast majority of
source sequences
within each type, and the vast majority of types (in analogy to the vast majority of
the parameter space in Rissanen's framework). In a way, 
this kind of a converse result
still contains some flavor of the probabilistic setting, because arguing
that the number of exceptional typical sequences is relatively small, is actually equivalent to imposing a
uniform distribution across the type and asserting that the induced probability of
violating the bound is small. A similar comment applies, of course, to the
exclusion of a minority of the types. The achievability result of \cite{WMF94}, on the other hand,
holds pointwise, for every sequence. 
A similar comment applies to \cite{me-univdis},
where asymptotically pointwise lossy compression was established with respect
to first order statistics (i.e., ``memoryless'' statistics) with an emphasis
on distortion-universality, similarly as in \cite{MW22a} and \cite{MW22b}.

A similar kind of a mix between the probabilistic setting and the
individual-sequence setting is adopted in this paper as well, in the
context of universal rate-distortion coding, but here, just like in
\cite{me-univsid}, there is no limitation
to finite-state encoders/decoders as in \cite{WMF94}. In particular, our converse theorem asserts
that given an arbitrary variable-rate code, and given an arbitrary distortion 
function within a certain wide class, the
majority of reproduction vectors that represent source sequences of a given
type (of any fixed order), must
have a code-length that is essentially at least as large as the negative
logarithm of the probability of a ball with normalized radius $D$ ($D$ being the allowed
per-letter distortion), centered at the given
source sequence. The probability of this ball is taken w.r.t.\ a universal distribution that is proportional to
$2^{-LZ(\hbx)}$, where $LZ(\hbx)$ is the code-length of LZ encoding of the
reproduction vector, $\hbx$. On the other hand, we also present a matching achievability
result, asserting that for every source sequence, this code length is
essentially achievable by random coding, using a universal ensemble of
codes, which is defined by
independent random selection, where each codeword is drawn under the
above-described universal probability distribution.

While the achievability result in \cite{me-univdis} was pointwise as well, it
was tailored to a memoryless structure in the sense that it was
given in terms of the rate-distortion function of the first-order empirical
distribution, which is blind to any empirical dependencies and repetitive
patterns within the source sequence. In this paper, we both extend the scope
to general individual sequences beyond the memoryless statistics and extend
the allowable class of distortion measures.
In terms of the technical aspects, the proof
of the achievablity result is very similar to the parallel proof in
\cite{me-univdis}, but the novelty lies considerably more in the converse theorem and its
proof.

The outline of this paper is as follows.
In Section \ref{ndb}, we establish the notation conventions, define a few
terms and quantities, and provide some background.
In Section \ref{converse}, we present the converse theorem and its proof.
In Section \ref{achievability}, we present the achievability theorem
and prove it. Finally, in Section \ref{discussion}, we summarize the paper and
discuss our results.

\section{Notation, Definitions and Background}
\label{ndb}

Throughout the paper, random variables will be denoted by capital
letters, specific values they may take will be denoted by the
corresponding lower case letters, and their alphabets
will be denoted by calligraphic letters. Random
vectors and their realizations will be denoted,
respectively, by capital letters and the corresponding lower case letters,
both in the bold face font. Their alphabets will be superscripted by their
dimensions. The source vector of length $n$, $(x_1,\ldots,x_n)$, with
components, $x_i$, $i=1,\ldots,n$, from a
finite-alphabet, $\calX$, will be denoted by $\bx$. The set of all such
$n$-vectors will be denoted by
$\calX^n$, which is the $n$--th order Cartesian power of $\calX$. 
Likewise, a reproduction vector of length $n$, $(\hx_1,\ldots,\hx_n)$, with 
components, $\hx_i$, $i=1,\ldots,n$, from a
finite-alphabet, $\hat{\calX}$, will be denoted by $\hbx\in\hat{\calX}^n$. 
We denote the cardinalities of $\calX$ and $\hat{\calX}$ by $J$ and $K$,
respectively.

For $i\le j$, the notation $x_i^j$ will be used to denote the substring
$(x_i,x_{i+1},\ldots,x_j)$.
Probability distributions will be denoted by the letter $P$ or $Q$ with
possible subscripts, depending on the context.
The probability of an event $\calE$ will be denoted by $\mbox{Pr}\{\calE\}$,
and the expectation
operator with respect to (w.r.t.) a probability distribution $P$ will be
denoted by
$\bE\{\cdot\}$.
For two positive sequences, $a_n$ and $b_n$, the notation $a_n\exe b_n$ will
stand for equality in the exponential scale, that is,
$\lim_{n\to\infty}\frac{1}{n}\log \frac{a_n}{b_n}=0$.
Similarly,
$a_n\lexe b_n$ means that
$\limsup_{n\to\infty}\frac{1}{n}\log \frac{a_n}{b_n}\le 0$, and so on.
The indicator function
of an event $\calE$ will be denoted by $\calI\{E\}$. The notation $[x]_+$
will stand for $\max\{0,x\}$.
The logarithmic function, $\log x$, will be understood to be defined to the
base 2. Logarithms to the base $e$ will be denote by $\ln$.

Let $\ell$ be a positive integer that divides $n$.
The $\ell$th order empirical distribution of $\bx\in\calX^n$, which will be
denoted by $\hat{P}_{\bx}^\ell$, is the vector of relative frequencies
$\{\hat{P}_{\bx}^\ell(a^\ell),~
a^\ell\in\calX^\ell\}$, where
\begin{equation}
\hat{P}_{\bx}^\ell(a^\ell)=\frac{\ell}{n}\sum_{i=0}^{n/\ell-1}\calI\{x_{i\ell+1}^{(i+1)\ell}=a^\ell\}.
\end{equation}
The set of all $\ell$th order empirical distributions of
sequences in $\calX^n$ will be denoted by $\calP_n^\ell$.
For $P^\ell\in\calP_n^\ell$, the type class,
$\{\bx\in\calX^n:~\hat{P}_{\bx}^\ell=P^\ell\}$, will be denoted by
$\calT_n(P^\ell)$. Likewise, $\calT_n(Q^\ell)$ will denote 
$\{\hbx\in\hat{\calX}^n:~\hat{P}_{\hbx}^\ell=Q^\ell\}$, where
$\hat{P}_{\hbx}^\ell$ is the $\ell$-th order empirical distribution of $\hbx$.
Finally, $\hat{P}_{\bx\hbx}^\ell$ will denote the
$\ell$th order joint
empirical distribution of $(\bx,\hbx)$, i.e.,
\begin{equation}
\hat{P}_{\bx\hbx}^\ell(a^\ell,b^\ell)=\frac{\ell}{n}\sum_{i=0}^{n/\ell-1}
\calI\{x_{i\ell+1}^{(i+1)\ell}=a^\ell,
\hx_{i\ell+1}^{(i+1)\ell}=b^\ell\},~~~(a^\ell,b^\ell)\in\calX^\ell\times\hat{\calX}^\ell.
\end{equation}

For a given positive integer $n$, a distortion function, $d$, is a function from
$\calX^n\times\hat{\calX}^n$ into $\reals^+$. In the two main parts of this paper,
different assumptions will be imposed on the distortion function.
\begin{enumerate}
\item For the achievability theorem, the distortion function can be completely
arbitrary. 
\item For the converse theorem, we assume that $d(\bx,\hbx)$ depends on $\bx$
and $\hbx$ only via their first order joint empirical distribution,
$\hat{P}_{\bx\hbx}^1$, and that for a given such distribution, it grows
linearly in $n$, that is, $d(\bx,\hbx)=n\rho(\hat{P}_{\bx\hbx}^1)$, where the
function $\rho$ is independent of $n$.
\end{enumerate}

Regarding item 2, additive distortion measures, which obviously comply with the
requirement, are given by linear functionals of $\hat{P}_{\bx\hbx}^1$.
However, here arbitrary non-linear functionals are allowed as well. 

A rate-distortion block code of length $n$ is a mapping,
$\phi_n:\calX^n\to\calB_n$, $\calB_n\subset\{0,1\}^*$, that maps the space of source vectors of length
$n$, $\calX^n$, into a set, $\calB_n$, of variable-length compressed bit
strings. The decoder is a mapping
$\psi_n:\calB^n\to\calC_n\subseteq\hat{\calX}^n$ that maps the set of
compressed variable-length binary strings into a reproduction codebook,
$\calC^n$. A block code
is called $d$-semifaithful if for every $\bx\in\calX^n$,
$d(\bx,\psi_n(\phi_n(\bx)))\le nD$. 
The code-length for $\bx$, denoted $L(\bx)$, is the number of bits of
$\phi_n(\bx)$. Since $L(\bx)$ depends on $\bx$ only via $\phi_n(\bx)$, we will
also denote it sometimes as $L(\phi_n(\bx))$ or by $L(\hbx)$ ($\hbx$ being the
reproduction vector pertaining to $\phi_n(\bx)$), with a slight abuse of notation.
For the
converse theorem, we assume that correspondence between
$\calB_n$ and $\calC_n$ is one-to-one. For the achievability theorem, we
consider prefix-free codes. Accordingly, the encoder can equivalently
be presented as a cascade of a reproduction encoder (a.k.a.\ vector
quantizer), which maps $\calX^n$ into $\calC_n$, followed by an entropy coder,
which maps $\calC_n$ into $\calB_n$ with no additional loss of information.

For the purpose of presenting both the converse theorem and the achievability theorem, we need to recall a few
terms and facts concerning the 1978 version of LZ algorithm (a.k.a.\ the LZ78
algorithm) \cite{ZL78}.
The incremental parsing procedure of the LZ78 algorithm is a procedure of
sequentially parsing a vector, $\hbx\in\hat{\calX}^n$, such that each new phrase is the shortest
string that has not been encountered before as a parsed phrase, with the
possible exception of the last phrase, which might be incomplete. For example,
the incremental parsing of the vector $\hbx=\mbox{abbabaabbaaabaa}$ is
$\mbox{a,b,ba,baa,bb,aa,ab,aa}$. Let $c(\hbx)$ denote the
number of phrases in $\hbx$ resulting from the incremental parsing procedure.
Let $LZ(\hbx)$ denote the
length of the LZ78 binary compressed code for $\hbx$. 
According to
\cite[Theorem 2]{ZL78},
\begin{eqnarray}
\label{lz-clogc}
LZ(\hbx)&\le&[c(\hbx)+1]\log\{2K[c(\hbx)+1]\}\nonumber\\
&=&c(\hbx)\log[c(\hbx)+1]+c(\hbx)\log(2J)+\log\{2K[c(\hbx)+1]\}\nonumber\\
&=&c(\hbx)\log c(\hbx)+c(\hbx)\log\left[1+\frac{1}{c(\hbx)}\right]+
c(\hbx)\log(2K)+\log\{2K[c(\hbx)+1]\}\nonumber\\
&\le&c(\hbx)\log c(\hbx)+\log
e+\frac{n(\log K)\log(2K)}{(1-\epsilon_n)\log
n}+\log[2K(n+1)]\nonumber\\
&\dfn&c(\hbx)\log c(\hbx)+n\cdot\epsilon(n),
\end{eqnarray}
where we remind that $K$ is the cardinality of $\hat{\calX}$, and where 
$\epsilon(n)$ clearly tends to zero as $n\to\infty$, at the rate of
$1/\log n$. 
We next define a {\em universal probability distribution} (see
also \cite{CM21}, \cite{MC20}):
\begin{equation}
\label{Udis}
U(\hbx)=\frac{2^{-LZ(\hbx)}}{\sum_{\hbx'\in\hat{\calX}^n}2^{-LZ(\hbx')}},~~~~\hbx\in\hat{\calX}^n.
\end{equation}
Finally, we define the $D$-sphere around $\bx$ as
\begin{equation}
\calS(\bx,D)=\{\hbx:~d(\bx,\hbx)\le nD\},
\end{equation}
and
\begin{equation}
U[\calS(\bx,D)]=\sum_{\hbx\in\calS(\bx,D)}U(\hbx).
\end{equation}
For later use, we also define
\begin{equation}
\hat{\calS}(\hbx,D)=\{\bx:~d(\bx,\hbx)\le nD\}.
\end{equation}

Our purpose is to derive upper and lower bounds on the smallest achievable
code length, $L(\bx)$, for $d$-semifaithful block codes of length $n$,
and individual sequences, $\{\bx\}$,
from a given $\ell$th order type class, $\calT_n(P^\ell)$. As
will be seen shortly, in
both the converse and the achievability theorems, the main term of the bound on the
length function will be $-\log(U[\calS(\bx,D)])$.

\section{The Converse Theorem}
\label{converse}

The following converse theorem asserts that even if the type class of the
source vector was known to the decoder ahead of time, the code length could
not be much smaller than $-\log(U[\calS(\bx,D)])$ for the vast majority of the
codewords pertaining to that type.

\begin{theorem}
\label{conversethm}
Let $\ell$ be a positive integer that divides $n$ and let $\hat{P}^\ell$ be
an arbitrary empirical distribution pertaining to a certain type class,
$\calT_n(\hat{P}^\ell)$, of source
sequences in $\calX^n$. Let $d$ be a distortion function that depends on
$(\bx,\hbx)$ only via $\hat{P}_{\bx\hbx}$.
Then, for every
every $d$-semifaithful variable-length block code, 
with one-to-one correspondence between $\calB_n$ and
$\calC_n$, and for every $\epsilon>0$, the following lower bound applies to a fraction of at least
$(1-2n^{-\epsilon})$ of the codewords,
$\{\phi_n(\bx),~\bx\in\calT_n(\hat{P}^\ell)\}$:
\begin{equation}
L(\phi_n(\bx))\ge -\log(U[\calS(\bx,D)])-n\Delta_n(\ell)-\epsilon\log n,
\end{equation}
where $\Delta_n(\ell)$ has the property $\lim_{n\to\infty}\Delta_n(\ell)=1/\ell$.
\end{theorem}

As a technical note, observe that $\Delta_n(\ell)$ can be made small only
when $\ell$ is chosen large, as $\Delta_n(\ell)$ behaves like $1/\ell$ for
fixed $\ell$ and large $n$. This suggests that the theorem is meaningful
mainly when $\ell$ is appreciably large, which is not surprising, because the
larger is $\ell$, the better one can exploit empirical dependencies within the
source sequence.

The remaining part of this section is devoted to the proof of Theorem
\ref{conversethm}.\\

\noindent
{\em Proof.}
We first establish a relationship that will be used later on.
For two given types $\calT_n(P^\ell)\subset\calX^n$ and
$\calT_n(Q^\ell)\subset\hat{\calX}^n$,
consider the quantity,
\begin{equation}
N(D)=\sum_{\bx,\hbx}\calI\{\bx\in\calT_n(P^\ell),~\hbx\in\calT_n(Q^\ell),~d(\bx,\hbx)\le
nD\}.
\end{equation}
We can evaluate $N(D)$ in two ways. The first is as follows:
\begin{eqnarray}
N(D)&=&\sum_{\bx\in\calT_n(P^\ell)}\bigg|T_n(Q^\ell)\bigcap\calS(\bx,D)\bigg|\\
&=&|\calT_n(P^\ell)|\cdot\bigg|T_n(Q^\ell)\bigcap\calS(\bx,D)\bigg|,
\end{eqnarray}
where the second equality is since $|T_n(Q^\ell)\bigcap\calS(\bx,D)\bigg|$ is the
same for all $\bx\in\calT_n(P^\ell)$, due to the permutation-invariance assumption on the distortion
function. By the same token, we can also express $N(D)$ in
the following manner:
\begin{eqnarray}
N(D)&=&\sum_{\hbx\in\calT_n(Q^\ell)}\bigg|T_n(P^\ell)\bigcap \hat{\calS}(\hbx,D)\bigg|\\
&=&|\calT_n(Q^\ell)|\cdot\bigg|T_n(P^\ell)\bigcap \hat{\calS}(\hbx,D)\bigg|,
\end{eqnarray}
which follows from the same consideration by symmetry. It follows then that
\begin{equation}
|\calT_n(P^\ell)|\cdot\bigg|T_n(Q^\ell)\bigcap
\calS(\bx,D)\bigg|=|\calT_n(Q^\ell)|\cdot\bigg|T_n(P^\ell)\bigcap
\hat{\calS}(\hbx,D)\bigg|,
\end{equation}
or, equivalently,
\begin{equation}
\frac{|T_n(P^\ell)|}{\bigg|T_n(P^\ell)\bigcap \hat{\calS}(\hbx,D)\bigg|}=
\frac{|T_n(Q^\ell)|}{\bigg|T_n(Q^\ell)\bigcap\calS(\bx,D)\bigg|}.
\end{equation}
Now, let $Q_*^\ell$ be the type of $\hbx$ that maximizes $|T_n(P^\ell)\bigcap
\hat{\calS}(\hbx,D)\bigg|$.
Then, the last equation implies that
\begin{equation}
\label{packing-rc}
\frac{|T_n(P^\ell)|}{\max_{\hbx\in\hat{\calX}^n}\bigg|T_n(P^\ell)\bigcap
\hat{\calS}(\hbx,D)\bigg|}=
\frac{|T_n(Q_*^\ell)|}{\bigg|T_n(Q_*^\ell)\bigcap
\calS(\bx,D)\bigg|},~~~\forall~\bx\in\calT_n(P^\ell).
\end{equation}
This relationship will be used shortly.

Let $P^\ell\in\calP_n^\ell$ be given. Any $d$-semifaithful code must fully cover
the type class $\calT_n(P^\ell)$ with spheres of radius $nD$ (henceforth, referred to as
$D$-spheres), centered at the various
codewords. Let
$\hbx_1,\ldots,\hbx_M\in\hat{\calX}^n$ be $M$ codewords. The number of members
of $\calT_n(P^\ell)$ that are covered by $\hbx_1,\ldots,\hbx_M\in\hat{\calX}^n$ is
upper bounded as follows.
\begin{eqnarray}
G&=&\bigg|\bigcup_{i=1}^M\left[\calT_n(P^\ell)\bigcap\hat{\calS}(\hbx_i,
D)\}\right]\bigg|\nonumber\\
&\le&\sum_{i=1}^M\bigg|\calT_n(P^\ell)\bigcap\hat{\calS}(\hbx_i,
D)\}\bigg|\nonumber\\
&\le&M\cdot\max_{\hbx\in\hat{\calX^n}}\bigg|\calT_n(P^\ell)\bigcap\hat{\calS}(\hbx,
D)\}\bigg|,
\end{eqnarray}
and so, the necessary condition for complete covering, which is
$G\ge|\calT_n(P^\ell)|$, amounts to
\begin{eqnarray}
M&\ge&\frac{|\calT_n(P^\ell)|}{\max_{\hbx\in\hat{\calX^n}}\bigg|\calT_n(P^\ell)\bigcap\hat{\calB}(\hbx,
D)\}\bigg|}\nonumber\\
&=&\frac{|T_n(Q_*^\ell)|}{\bigg|T_n(Q_*^\ell)\bigcap
\calS(\bx,D)\bigg|}\nonumber\\
&\dfn&M_0,
\end{eqnarray}
where the second line is by (\ref{packing-rc}).
Consider now a variable-length code with a codebook of size $M$.
Let $L(\hbx)$ denote the length (in bits) of the compressed binary string
that represents $\hbx$. The number of codewords with
$L(\hbx)\le \log M-\epsilon\log n$ is upper bounded as follows:
\begin{eqnarray}
|\{\hbx\in\calC_n:~L(\hbx)\le \log M-\epsilon\log n\}|&=&\sum_{k=1}^{\log
M-\epsilon\log
n}|\{\hbx:~L(\hbx)=k\}|\nonumber\\
&\le&\sum_{k=1}^{\log M-\epsilon\log
n} 2^k\nonumber\\
&=&2^{\log M-\epsilon\log n+1}-1\nonumber\\
&<&2n^{-\epsilon} M,
\end{eqnarray}
where in the first inequality we have used the assumed one-to-one property of
the mapping between the reproduction codewords and their variable-length
compressed binary representations.
It follows then that for at least $M(1-2n^{-\epsilon})$ out of the $M$
codewords in $\calC^n$ (that is, the vast majority codewords), we have
\begin{eqnarray}
L(\phi_n(\bx))&\ge&\log M-\epsilon\log n\nonumber\\
&\ge&\log M_0-\epsilon\log n\nonumber\\
&=&-\log\left(\frac{\bigg|T_n(Q_*^\ell)\bigcap
\calS(\bx,D)\bigg|}{|T_n(Q_*^\ell)|}\right)
-\epsilon \log n\nonumber\\
&=&-\log\left[\sum_{\hbx\in\calS(\bx,D)}U_{Q_*}(\hbx)\right]-\epsilon\log n,
\end{eqnarray}
where $U_{Q_*}$ is the uniform probability distribution across the type
class $Q_*^\ell$, i.e.,
\begin{equation}
U_{Q_*}(\hbx)=\left\{\begin{array}{ll}
\frac{1}{|\calT_n(Q_*^\ell)|} & \hbx\in\calT_n(Q_*^\ell)\\
0 & \mbox{elsewhere}\end{array}\right.
\end{equation}
We now argue that for every $\hbx\in\hat{\calX}^n$
\begin{equation}
\label{UQ2LZ}
U_{Q_*}(\hbx)\le\exp_2\{-LZ(\hbx)
+n\Delta_n(\ell)\}.
\end{equation}
For $\hbx\notin\calT_n(Q_*^\ell)$, this is trivial as the l.h.s.\ is equal to zero.
For $\hbx\in\calT_n(Q_*^\ell)$, we have the following consideration:
Combining eqs.\ (30) and
(32) of \cite{me20} together with the inequality \cite[p.\ 17, Lemma
2.3]{CK11},
\begin{equation}
|\calT_n(Q_*^\ell)|\ge \left(\frac{n}{\ell}+1\right)^{-K^\ell}\cdot
2^{nH_{Q_*}(\hat{X}^\ell)/\ell},
\end{equation}
where
\begin{equation}
H_{Q_*}(\hat{X}^\ell)=-\sum_{b^\ell\in\hat{\calX}^\ell}Q_*^\ell(b^\ell)\log
Q_*^\ell(b^\ell),
\end{equation}
we have
\begin{eqnarray}
\label{logT}
\log|\calT_n(Q_*^\ell)|&\ge&c(\hbx)\log
c(\hbx)-n\delta_n(\ell)-\frac{K^\ell}{n}\log\left(\frac{n}{\ell}+1\right)\nonumber\\
&\ge&LZ(\hbx)-n\epsilon(n)
-n\delta_n(\ell)-\frac{K^\ell}{n}\log\left(\frac{n}{\ell}+1\right)\nonumber\\
&\dfn&LZ(\hbx)-n\Delta_n(\ell),
\end{eqnarray}
where
\begin{equation}
\delta_n(\ell)=\frac{\log[4S^2(\ell)]\log K}{(1-\epsilon_n)\log n}
+\frac{S^2(\ell)\log[4S^2(\ell)]}{n}+\frac{K^\ell}{n}\log\left(\frac{n}{\ell}+1\right)+\frac{1}{\ell},
\end{equation}
and
\begin{equation}
S(\ell)=\frac{J^{\ell+1}-1}{J-1},
\end{equation}
and where the second inequality in (\ref{logT}) follows from (\ref{lz-clogc}).
The last line of (\ref{logT}) is equivalent to (\ref{UQ2LZ}).
It follows then that
for at least $M(1-2\cdot n^{-\epsilon})$ out of the $M$
codewords in $\calC^n$,
\begin{eqnarray}
L(\phi_n(\bx))&\ge&-\log\left[\sum_{\hbx\in\calS(\bx,D)}2^{-LZ(\hbx)}
\right]-n\Delta_n(\ell)-\epsilon\log n\nonumber\\
&=&-\log\left[\sum_{\hbx\in\calS(\bx,D)}\frac{2^{-LZ(\hbx)}}
{\sum_{\hbx'\in\hat{\calX}^n}2^{-LZ(\hbx')}}
\right]-\nonumber\\
& &\log\left(\sum_{\hbx\in\hat{\calX}^n}2^{-LZ(\hbx)}
\right)-n\Delta_n(\ell)-\epsilon\log n\nonumber\\
&\ge&-\log(U[\calS(\bx,D)])
-n\Delta_n(\ell)-\epsilon\log n,
\end{eqnarray}
where in the last step we have applied Kraft's inequality to
the LZ code-length function. This completes the proof of Theorem
\ref{conversethm}.

\section{The Achievability Theorem} 
\label{achievability}

The lower bound of Theorem \ref{conversethm} naturally
suggests achievability using the universal distribution, $U$, for random
selection of the various codewords. The basic idea is quite standard and
simple: The quantity $U[\calS(\bx,D)]$ is the
probability that a single randomly chosen reproduction vector, drawn
under $U$, would fall within
distance $nD$ from the source vector, $\bx$. If all reproduction vectors
are drawn independently under $U$, then the typical number of such random selections
that it takes before one sees the first one in $\calS(\bx,D)$, is of the
exponential order of $1/U[\calS(\bx,D)]$. Given that the codebook is revealed
to both the encoder and decoder, once it has been selected, the encoder merely
needs to transmit
the index of the first such reproduction vector within the
codebook, and the description length of that index can be made essentially as small as
$\log\{1/U[\calS(\bx,D)]\}=-\log(U[\calS(\bx,D))$.
We use this simple idea to prove achievability for an
arbitrary distortion measure.
The proof is very similar to the parallel proof in
\cite{me-univdis}, and it is presented here mainly for completeness.

The achievability theorem is the following.
\begin{theorem}
\label{thm2}
Let $d:\calX^n\times\hat{\calX}^n\to\reals^+$ be an arbitrary distortion function.
Then, for every $\epsilon>0$, there
exists a sequence of $d$-semifaithful, variable-length block codes of block length $n$,
such that for every
$\bx\in\calX^n$, the code length for $\bx$ is upper bounded by
\begin{equation}
L(\bx)\le-\log(U[\calS(\bx,D)])+(2+\epsilon)\log n +c+\delta_n,
\end{equation}
where $c> 0$ is a constant and $\delta_n=O(nJ^ne^{-n^{1+\epsilon}})$.
\end{theorem}

\noindent
{\em Proof.}
The proof is based on the
following simple well known fact: 
Given a source vector $\bx\in\calX^n$ and a codebook, $\calC_n$, let $I(\bx)$
denote the index, $i$, of the first vector, $\hbx_i$, such that $d(\bx,\hat{\bx}_i)\le nD$, namely,
$\hbx_i\in\calS(\bx,D)$.
If all reproduction vectors are drawn independently under $U$, 
then, for every positive integer, $N$:
\begin{equation}
\label{fact1}
\mbox{Pr}\{I(\bx)> N\}=(1-U[\calS(\bx,D)])^N
=\exp\{N\ln(1-U[\calS(\bx,D)]\}\le \exp\{-N\cdot U[\calS(\bx,D)]\},
\end{equation}
and so, if $N=N_n= e^{\lambda_n}/U[\calS(\bx,D)]$,
for some arbitrary positive sequence, $\{\lambda_n\}$, that tends to infinity,
then
\begin{equation}
\label{fact2}
\mbox{Pr}\{I(\bx)> N_n\}\le \exp\{-e^{\lambda_n}\}.
\end{equation}
This fact will be used few times in this section.

For later use, we also need the following uniform lower bound to
$U[\calS(\bx,D)]$: For a given $\bx$, let $\hbx_0\in\hat{\calX}^n$
denote an arbitrary
reproduction vector within $\calS(\bx,D)$. Then,
\begin{eqnarray}
\label{ulb}
U[\calS(\bx,D)]&\ge&U(\hbx_0)\\
&=&\frac{2^{-LZ(\hbx_0)}}{\sum_{\hbx\in\hat{\calX}^n}2^{-LZ(\hbx)}}\\
&\ge&2^{-LZ(\hbx_0)}.
\end{eqnarray}
Next, observe that
$LZ(\hbx_0)$ is maximized by the $K$-ary extension of the counting sequence
\cite[p.\ 532]{ZL78}, which is defined as follows: For $i=1,2,\ldots,m$ ($m$
-- positive integer), let
$u(i)$ denote the $K$-ary string of length $iK^i$ that lists, say, in lexicographic
order, all the $K^i$ words from $\hat{\calX}^i$, and let
$\hbx_0=(u(1)u(2)\ldots u(m))$, whose length is
\begin{eqnarray}
n&=&\sum_{i=1}^miK^i\nonumber\\
&=&K\cdot\sum_{i=1}^miK^{i-1}\nonumber\\
&=&K\cdot\frac{\partial}{\partial
K}\left(\sum_{i=1}^mK^i\right)\nonumber\\
&=&K\cdot\frac{\partial}{\partial
K}\left(\frac{K^{m+1}-K}{K-1}\right)\nonumber\\
&=&\frac{K}{(K-1)^2}[mK^{m+1}-(m+1)K^m+1].
\end{eqnarray}
The LZ incremental parsing of $\hbx_0$, which is exactly
$(u(1),u(2),\ldots,u(m))$, yields:
\begin{equation}
c(\hbx_0)=\sum_{i=1}^mK^i=\frac{K^{m+1}-K}{K-1},
\end{equation}
and so, considering eq.\ (\ref{lz-clogc}), it follows that
$LZ(\hbx_0)\le(1+\epsilon_n)n\log K$ for some $\epsilon_n\to 0$ as
$n\to\infty$.\footnote{As an alternative to this upper bound on the LZ code
length, one can slightly modify the LZ algorithm as follows: If $LZ(\hbx)\le
n\log K$ use the LZ algorithm as usual, otherwise,
send $\hbx$ uncompressed using $n\log K$ bits. To distinguish between the two
modes of operation, append a flag bit to
indicate whether or not the data is LZ-compressed. The modified code-length would then be
$LZ'(\hbx)=\min\{LZ(\hbx),n\log K\}+1$. Now, replace $LZ(\hbx)$ by $LZ'(\hbx)$
in all places throughout this paper, including the definition of $U$.}
It follows then that
\begin{equation}
\label{ulb1}
U[\calS(\bx,D)]\ge 2^{-n(1+\epsilon_n)\log K}.
\end{equation}

Consider now an independent random selection of all reproduction vectors to
form a codebook, $\calC_n$, of size $M=A^n$ ($A>K$) codewords,
$\hbx_1,\hbx_2,\ldots,\hbx_M$, according to $U$.
Once the codebook $\calC_n=\{\hbx_1,\hbx_2,\ldots,\hbx_M\}$ has been drawn,
it is revealed to both the encoder and the decoder.
Consider next the following encoder.
As defined before, let $I(\bx)$ be defined as the
index of the first
codeword that falls within $\calS(\bx,D)$, but now, with
the small modification that
if none of the $A^n$ codewords fall in $\calS(\bx,D)$, then
we define $I(\bx)= A^n$ nevertheless (and then the encoding fails).
Next, we define the following probability distribution over the positive integers,
$1,2,\ldots,A^n$:
\begin{equation}
u[i]=\frac{1/i}{\sum_{k=1}^{A^n}1/k},~~~~i=1,2,\ldots,A^n.
\end{equation}
Given $\bx$, the encoder finds $I(\bx)$ and
encodes it using a variable-rate
lossless code with the length function (in bits, and ignoring the
integer length
constraint),
\begin{eqnarray}
\label{Ld}
L(\bx)&=&-\log u[I(\bx)]\nonumber\\
&\le&\log I(\bx)+\log\left(\sum_{k=1}^{A^n}\frac{1}{k}\right)\nonumber\\
&\le&\log I(\bx)+\log(\ln A^n+1)\nonumber\\
&=&\log I(\bx)+\log(n\ln A+1)\nonumber\\
&\le&\log I(\bx)+\log n +c,
\end{eqnarray}
where $c=\log(\ln A+1)$.
It follows that the expected codeword length for $\bx\in\calX^n$ (w.r.t.\ the
randomness of the code) is upper bounded by:
\begin{eqnarray}
\label{basicub}
\bE\{L(\bx)\}&\le&\bE\{\log I(\bx)\}+\log n+c\nonumber\\
&\le&\log \bE\{I(\bx)\}+\log n+c\nonumber\\
&=&\log \left(\sum_{k=1}^{A^n}k\cdot\left(1-U[\calS(\bx,D)]
\right)^{k-1}\cdot U[\calS(\bx,D)]
+A^n\cdot\left(1-U[\calS(\bx,D)]\right)^{A^n}\right)+\log
n+c\nonumber\\
&=&\log \left(\sum_{k=1}^{\infty}\min\{k,A^n\}\cdot\left(1-U[\calS(\bx,D)]
\right)^{k-1}\cdot U[\calS(\bx,D)]
\right)+\log n+c\nonumber\\
&\le&\log\left\{\sum_{k=1}^\infty k\cdot\left(1-U[\calS(\bx,D)]
\right)^{k-1}\cdot U[\calS(\bx,D)]\right\}+\log
n+c\nonumber\\
&=&\log\left(\frac{1}{U[\calS(\bx,D)]}\right)+\log n+c,
\end{eqnarray}
and we denote
\begin{equation}
\label{Ldplus}
L^+(\bx)\dfn\log\left(\frac{1}{U[\calS(\bx,D)]}\right)+\log n+c.
\end{equation}
Consider now the quantity
\begin{eqnarray}
E_n&\dfn&\bE\bigg\{\max\bigg(\max_{\bx\in\calX^n}\calI\{d(\bx,\hat{\bX})>
nD\},\nonumber\\
& &\left[\max_{\bx\in\calX^n}\left(L(\bx)-L^+(\bx)-(1+\epsilon)\log
n\right)\right]_+\bigg)\bigg\},
\end{eqnarray}
where the expectation is w.r.t.\ the randomness of the
code, $\calC_n$.
If $E_n$ can be upper bounded by $\delta_n$, which tends to zero
as $n\to\infty$, this will imply that there must exist a
code for which both
\begin{equation}
\label{dist}
\max_{\bx\in\calX^n}\calI\{d(\bx,\hbx)>
nD\}\le\delta_n
\end{equation}
and
\begin{equation}
\label{length}
\max_{\bx\in\calX^n}\left(L(\bx)-
L^+(\bx)-(1+\epsilon)\log
n\right)\le\delta_n
\end{equation}
at the same time.
Observe that since the left-hand side of (\ref{dist}) is either zero or one,
then if we know that
it must be less than $\delta_n\to 0$, for some codebook, $\calC_n$, it means
that it must vanish
as soon as $n$ is large enough such that $\delta_n < 1$, namely,
$d(\bx,\hbx)\le nD$ for all $\bx$, in other words, the code is
$d$-semifaithful.
Also, by (\ref{length}), for the same codebook, we must have
\begin{equation}
L(\bx)\le L^+(\bx)+(1+\epsilon)\log
n+\delta_n~~~~\bx\in\calX^n,
\end{equation}
and $\delta_n$ adds a negligible redundancy term.

To prove that $E_n\to 0$, we first use the simple fact
that the maximum of two non-negative numbers is upper bounded by their
sum, i.e.,
\begin{eqnarray}
\label{sum}
E_n&\le&\bE\left\{\max_{\bx\in\calX^n}\calI\{d(\bx,\hat{\bX})>
nD\}\right\}+\nonumber\\
& &\bE\left\{\left[\max_{\bx\in\calX^n}\left(L(\bx)-
L^+(\bx)-(1+\epsilon)\log
n)\right)\right]_+\right\},
\end{eqnarray}
and therefore, it is sufficient to prove that each one of these terms tends to
zero. As for the first term, we have:
\begin{eqnarray}
\bE\left\{\max_{\bx\in\calX^n}\calI\{d(\bx,\hat{\bX})>
nD\}\right\}&\le&
\bE\left\{\sum_{\bx\in\calX^n}\calI\{d(\bx,\hat{\bX})>
nD\}\right\}\nonumber\\
&=&\sum_{\bx\in\calX^n}\bE\left\{\calI\{d(\bx,\hat{\bX})>
nD\}\right\}\nonumber\\
&=&\sum_{\bx\in\calX^n}\mbox{Pr}\{d(\bx,\hat{\bX})>
nD\}\nonumber\\
&=&\sum_{\bx\in\calX^n}\left(1-U[\calS(\bx,D)]
\right)^{A^n}\nonumber\\
&\le&\sum_{\bx\in\calX^n}\exp\left\{-A^nU[\calS(\bx,D)]
\right\}\nonumber\\
&\lea&\sum_{\bx\in\calX_n}\exp\left\{-\exp\left\{n\left[\ln
A-(1+\epsilon_n)\ln K
\right]\right\}\right\}\nonumber\\
&\le&J^n\exp\left(-\exp\left\{n\left[\ln A-(1+\epsilon_n)\ln
K\right]\right\}\right\},
\end{eqnarray}
where in (a) we have used (\ref{ulb1}).
This quantity decays
double-exponentially rapidly as $n\to\infty$ since we have assumed
$A > K$.

As for the second term of (\ref{sum}), we have:
\begin{eqnarray}
& &\bE\left\{\left[\max_{\bx\in\calX^n}\left(L(\bx)-
L^+(\bx)-(1+\epsilon)\log
n\right)\right]_+\right\}\nonumber\\
&\lea&\bE\left\{\left[\max_{\bx\in\calX^n}\left(\log
I(\bx)-\log\frac{1}{U[\calS(\bx,D)]}-(1+\epsilon)\log n
\right)\right]_+\right\}\nonumber\\
&=&\int_0^\infty\mbox{Pr}\left\{\max_{\bx\in\calX^n}\left[\log
I(\bx)-\log\frac{1}{U[\calS(\bx,D)]}-
(1+\epsilon)\log
n)\right]\ge s\right\}\mbox{d}s\nonumber\\
&=&\int_0^{n\log A}
\mbox{Pr}\left\{\max_{\bx\in\calX^n}\left[\log
I(\bx)-\log\frac{1}{U[\calS(\bx,D)]}-(1+\epsilon)\log n\right]
\ge s\right\}\mbox{d}s\nonumber\\
&=&\int_0^{n\log A}
\mbox{Pr}\left[\bigcup_{\bx\in\calX^n}\left\{I(\bx)\ge
\frac{2^{(1+\epsilon)\log
n+s}}{U[\calS(\bx,D)]}\right\}\right]\mbox{d}s\nonumber\\
&\le&\sum_{\bx\in\calX^n}\int_0^{n\log A}
\mbox{Pr}\left\{I(\bx)\ge\frac{2^{(1+\epsilon)\log
n+s}}{U[\calS(\bx,D)]}
\right\}\mbox{d}s\nonumber\\
&\leb&\sum_{\bx\in\calX^n}\int_0^{n\log A}
\exp\{-2^sn^{1+\epsilon}\}
\mbox{d}s\nonumber\\
&\le&J^n\cdot(n\log
A)\cdot\exp\{-n^{1+\epsilon}\},
\end{eqnarray}
where in (a) we have used (\ref{Ld}) and (\ref{Ldplus}), and in
(b) we have used (\ref{fact2}). The right-most side of this chain of
inequalities clearly
decays as well when $n$ grows without bound. 
This completes the proof.

\section{Summary and Discussion}
\label{discussion}

By deriving asymptotically matching upper and lower bounds, we have established the quantity
$-\frac{1}{n}\log(U[\calS(\bx,D)])$ as having the significance of an
empirical rate distortion function for individual sequences. While this
quantity is not easy to calculate for large $n$, the operative meaning of our
results is that we propose a universal ensemble for rate-distortion coding.
According to this ensemble, the codewords are drawn independently under the
probability distribution that is proportional to $2^{-LZ(\hbx)}$.

There are several
observations, insights and perspectives that should be addressed.\\

\noindent
{\em Relation to earlier converse bounds}. 
The converse bound is given in terms of the probability of a sphere of
radius $nD$ around the source vector $\bx$, under the universal distribution, $U$,
defined in (\ref{Udis}). This is intimately related to a converse result due
to Kontoyiannis and Zhang \cite[Theorem 1, part i)]{KZ02}, which states that
for any $d$-semifaithful code, there exists a probability distribution $Q$ on
$\hat{\calX}^n$ such that $L(\bx)\ge -\log(Q[\calS(\bx,D)])$ for all $\bx$
(see also
\cite{Kontoyiannis00}). Here, upon giving up any claims on a minority of the
codewords pertaining
to a given type class, we derived a lower bound of essentially the same form with the
benefit of specifying a concrete choice of the distribution $Q$,
i.e., we propose
$Q=U$, the universal distribution (unlike the distribution in \cite[Section
III.A]{KZ02}, which is proportional to $2^{-L(\hbx)}$ across the codebook).\\

\noindent
{\em Interpretation of the main term of the bound.} Since $LZ(\hbx)$ is
essentially bounded by a linear function of $n$ (see (\ref{ulb1})), we can approximate the main
term as follows:
\begin{eqnarray}
\label{rd}
-\log(U[\calS(\bx,D)])&\le&-\log\left(\sum_{\hbx\in\calS(\bx,D)}2^{-LZ(\hbx)}\right)\nonumber\\
&=&-\log\left(\sum_{L\ge
1}2^{-L}\cdot\bigg|\{\hbx:~LZ(\hbx)=L\}\bigcap\calS(\bx,D)\bigg|\right)\nonumber\\
&\approx&\min_{L\ge
1}\left\{L-\log\bigg|\{\hbx:~LZ(\hbx)=L\}\bigcap\calS(\bx,D)\bigg|\right\}.
\end{eqnarray}
This expression, when normalized by $n$, can be viewed as a certain extension of the 
rate distortion function, from the memoryless case to the general case, in the
following sense: For a memoryless source $P$, the rate-distortion function has
the following representation, which is parallel to (\ref{rd}):
\begin{equation}
R(D)=\min_{P_{\hX}}\left[H(\hX)-\max_{\{P_{X|\hX}:~\bE d(X,\hX)\le
D,~P_X=P\}}H(\hX|X)\right],
\end{equation}
where the maximum over the empty set is understood to be
$-\infty$. Indeed, if we replace $U$ by the the uniform distribution across
the first-order type
pertaining to the optimal $P_{\hX}$, this is the corresponding single-letter
expression of $-\log(P_{\hX}[\calS(\bx,D)])$ that is obtained using the method of types \cite{CK11}.\\

\noindent
{\em Comparing to the LZ description length of the most compressible
$\hbx\in\calS(\bx,D)$.}
Since our achievable bound involves LZ compression, it is interesting to
compare it to the conceptually simple coding scheme that encodes $\bx$ by the
vector $\hbx$ that minimizes $LZ(\hbx)$ within $\calS(\bx,D)$. Consider the
following chain of equalities and inequalities:
\begin{eqnarray}
\min_{\hbx\in\calS(\bx,D)}LZ(\hbx)&=&-\log\left(\max_{\hbx\in\calS(\bx,D)}
2^{-LZ(\hbx)}\right)\nonumber\\
&\ge&-\log\left(\sum_{\hbx\in\calS(\bx,D)}
2^{-LZ(\hbx)}\right)\nonumber\\
&\ge&-\log\left(\sum_{\hbx\in\calS(\bx,D)}
\frac{2^{-LZ(\hbx)}}{\sum_{\hbx'\in\hat{\calX}^n}2^{-LZ(\hbx')}}\right)\nonumber\\
&=&-\log(U[\calS(\bx,D)]),
\end{eqnarray}
which means that the performance of our proposed scheme 
is never worse (and conceivably, often much better) than that of selecting the vector $\hbx$ with the smallest
$LZ(\hbx)$ among all reproduction vectors in $\calS(\bx,D)$. The reason for
the superiority of the proposed scheme is that it takes advantage of the fact
that $\hbx$ cannot be any vector in $\hat{\calX}^n$, but it must be a member
of the codebook, $\calC_n$, i.e., one of the possible outputs of a vector
quantizer. On the other hand, in view of \cite{ZL78}, $\min_{\hbx\in\calS(\bx,D)}LZ(\hbx)$ 
is essentially achievable upon compressing the output of a certain
reproduction encoder (or vector quantizer) using a finite--state
encoder, but a finite-state machine does not have enough memory resources to take
advantage of the fact that vectors
outside $\calC_n$ cannot be encountered by the encoder. Another interesting comparison between the two schemes
is in terms of computational complexity. While in our scheme, the encoder has
to carry out typically about $1/U[\calS(\bx,D)]$ distortion calculations before finding 
the first $\hbx\in\calS(\bx,D)$, in the alternative scheme the number of
calculations is $|\calS(\bx,D)|$. The former is decreasing function of $D$,
whereas the latter is an increasing function of $D$. Therefore, in terms of
computational complexity, the preference between the two schemes might depend
on $D$. Specifically, for an additive distortion measure, it is easy to see
that 
\begin{equation}
\frac{1}{U[\calS(\bx,D)]}\lexe \exp_2\{nR(D,P_{\bx}^1)\} 
\end{equation}
and, by the method of types \cite{CK11}:
\begin{equation}
|\calS(\bx,D)|\exe \exp_2\{nE(D,P_{\bx}^1))\}\dfn \exp_2[\max\{H(\hX|X),~\bE d(X,\hX)\le D,~P_X=P_{\bx}^1\}].
\end{equation}
Therefore, whenever $D$ is large enough such that $R(D,P_{\bx}^1)< E(D,P_{\bx}^1))$, it is guaranteed
that the coding scheme proposed here is computationally less demanding than the
alternative scheme of minimizing $LZ(\hbx)$ across $\calS(\bx,D)$.\\

\noindent
{\em Implementation of the random coding distribution.}
The universal random coding distribution is not difficult to implement. One
way to do this is by
feeding the LZ decoder with a sequence of purely random bits (fair coin
tosses) until we have obtained $n$
symbols at the decoder output. The details can be found in \cite{MC20}.\\

\noindent
{\em Universality w.r.t.\ the distortion measure.} 
As mentioned in the Introduction, in \cite{MW22a}, \cite{MW22b} and \cite{me-univdis}, there are results on the
existence of rate-distortion codes that are universal, not only in terms of
the source, but also in the sense of the distortion measure. Since the proof
of our achievability scheme is very similar to that of \cite{me-univdis}, it
is possible to extend the achievability proof here too, so as to make our code
distortion-universal for a wide class of distortion measures. This can be
carried out by redefining $E_n$ to include maximization of both terms over a
dense grid of distortion functions, as was done in \cite{me-univdis}. We opted
not to include this in the present paper since it is straightforward, given the
results we already have here and in \cite{me-univdis}.\\



\newpage
\begin{thebibliography}{AA}

\bibitem{AM98}
E.~Arikan and N.~Merhav, ``Guessing subject to distortion,''
{\em IEEE Trans.\ Inform.\ Theory}, vol.\ 44,
no.\ 3, pp.\ 1041--1056, May 1998.

\bibitem{Berger71}
T.~Berger, {\em Rate Distortion Theory - A Mathematical Basis for Data
Compression}, Prentice-Hall Inc., Englewood Cliffs, N.J., 1971.

\bibitem{CM21}
A.~Cohen and N.~Merhav, ``Universal randomized guessing subjected to
distortion,'' {\em IEEE Trans.\ Inform.\ Theory}, vol.\ 68, no.\ 12,
pp.\ 7714--7734, December 2022.

\bibitem{CT06}
T.~M.~Cover and J.~A.~Thomas, {\em Elements of Information Theory},
John Wiley \& Sons, Hoboken N.~J., 2006.

\bibitem{CK11}
I.~Csisz\'ar and J.~K\"orner, {\em Information Theory - Coding Theorems for
Discrete Memoryless Systems}, Second Edition, Cambridge University Press,
Cambridge, UK, 2011.

\bibitem{Davisson73}
L.~D.~Davisson, `Universal noiseless coding,'' 
{\it IEEE Trans.\ Inform.\ Theory}, vol.\
IT--29, no.\ 6, pp.\ 783--795, November 1973.

\bibitem{Gallager68}
R.~G.~Gallager, {\em Information Theory and Reliable Communication},
John Wiley \& Sons, New York 1968.

\bibitem{Gallager76}
R.~G.~Gallager, ``Source coding with side information and universal 
coding,'' LIDS-P-937, M.I.T., 1976.

\bibitem{Gray90}
R.~M.~Gray, {\em Source Coding Theory}, Kluwer Academic Publishers, Boston,
1990.

\bibitem{Kontoyiannis00}
I.~Kontoyiannis, ``Pointwise redundancy in lossy data compression and
universal lossy data compression,''
{\em IEEE Trans.\ Inform.\ Theory}, vol.\ 46, no.\ 1, pp.\ 136-152, January
2000.

\bibitem{KZ02}
I.~Kontoyiannis and J.~Zhang, ``Arbitrary source models and Bayesian codebooks
in rate-distortion theory,'' {\em IEEE Trans.\ Inform.\ Theory}, vol.\ 48,
no.\ 8, pp.\ 2276--2290, August 2002.

\bibitem{MW22a}
A.~Mahmood and A.~B.~Wagner, ``Lossy compression with universal
distortion,''\\
{\tt https://arxiv.org/pdf/2110.07022.pdf} February 9, 2022.

\bibitem{MW22b}
A.~Mahmood and A.~B.~Wagner, ``Minimax rate-distortion,''\\
{\tt https://arxiv.org/pdf/2202.04481.pdf} February 9, 2022.

\bibitem{me-univdis}
N.~Merhav, ``$D$-semifaithful codes that are universal over both memoryless sources and
distortion measures,'' submitted for publication. Also, available on-line at:
{\tt http://arxiv.org/pdf/2203.03305.pdf}

\bibitem{me93}
N.~Merhav, ``A comment on `A rate of convergence result for
a universal $d-$semifaithful code',''
{\em IEEE Trans.\ Inform.\ Theory}, vol.\ 41, no.\ 4, pp.\ 1200-1202, July
1995.

\bibitem{me14}
N.~Merhav, ``On the data processing theorem in the semi-deterministic
setting,'' {\em IEEE Trans.\ Inform.\ Theory}, vol.\ 60, no.\ 10, pp.\ 6032--6040, October
2014.

\bibitem{me20}
N.~Merhav, ``Guessing individual sequences: generating randomized
guesses using finite-state machines,'' {\em IEEE Trans.\ Inform.\ Theory},
vol.\ 66, no.\ 5, pp.\ 2912--2920, May 2020.

\bibitem{me21}
N.~Merhav, ``Encoding individual source sequences for the wiretap channel,''
{\em Entropy}, 23(12) 1694, December 17, 2021.

\bibitem{me22}
N.~Merhav, ``Finite-state source-channel coding for individual source
sequences with source side information at the decoder,'' {\em IEEE
Trans.\ Inform.\ Theory}, 
vol.\ 68, no.\ 3, pp.\ 1532--1544, March 2022.

\bibitem{MC20}
N.~Merhav and A.~Cohen, ``Universal randomized guessing with
application to asynchronous decentralized brute--force attacks,''
{\it IEEE Trans.\ Inform.\ Theory}, vol.\ 66, no.\ 1, pp.\ 114--129, January
2020.

\bibitem{MF95}
N.\ Merhav and M.\ Feder, ``A strong version of
the redundancy--capacity theorem of universal coding,''
{\em IEEE Trans. Inform. Theory}, vol.\ 41, no.\ 3, pp.\ 714-722, May 1995.

\bibitem{MZ06}
N.~Merhav and J.~Ziv, ``On the Wyner-Ziv problem for individual sequences,''
{\em IEEE Trans.\ Inform.\ Theory}, vol.\ 52, no.\ 3, pp. 867--873, March 2006.

\bibitem{WMF94}
M.~J.~Weinberger, N.~Merhav, and M.~Feder, ``Optimal sequential probability
assignment for individual sequences,'' {\em IEEE Trans.\ Inform.\ Theory},
vol.\ 40, no.\ 2, pp.\ 384--396, March 1994.

\bibitem{OS90}
D.~S.~Ornstein and P.~C.~Shields, ``Universal almost sure data compression,''
{\em Ann.\ Probab.}, vol.\ 18, no.\ 2, pp.\ 441--452, 1990.

\bibitem{Rissanen84}
J.~Rissanen, ``Universal coding, information, prediction, and 
estimation,'' {\em IEEE Transactions on Information Theory\/},
vol.~IT--30, no.~4, pp.~629--636, July 1984.

\bibitem{SP21}
J.~F.~Silva and P.~Piantanida,
``On universal $d$-semifaithful coding for
memoryless sources with infinite alphabets,''
{\em IEEE Transactions on Information Theory\/},
vol.~68, no.~4, pp.~2782--2800, April 2022.

\bibitem{VO79}
A.~J.~Viterbi and J.~K.~Omura, {\em Principles of Digital Communication and
Coding}, McGraw-Hill Inc., New York, 1979.

\bibitem{YS93}
B.~Yu and T.~Speed, ``A rate of convergence result for a universal
$d$-semifaithful code,'' {\em IEEE Trans.\ Inform.\ Theory}, vol.\ 39, no.\ 3,
pp.\ 813--820, May 1993.

\bibitem{ZYW97}
Z.~Zhang, E.-h.~Yang, and V.~Wei, ``The redundancy of source coding with a
fidelity criterion. I. known statistics,'' {\em IEEE Trans.\ Inform.\ Theory},
vol.\ 43, no.\ 1, pp.\ 71--91, January 1997.

\bibitem{Ziv78}
J.~Ziv, ``Coding theorems for individual sequences,'' {\em IEEE Trans.\
Inform.\ Theory},
vol.\ IT--24, no.\ 4, pp.\ 405--412, July 1978.

\bibitem{Ziv80}
J.~Ziv, ``Distortion-rate theory for individual sequences,'' 
{\em IEEE Trans.~Inform.~Theory\/},
vol.~IT--26, no.~2, pp.~137--143, March 1980.

\bibitem{Ziv84}
J.~Ziv, ``Fixed-rate encoding of individual sequences with side 
information,'' {\em IEEE Transactions on Information Theory\/},
vol.~IT--30, no.~2, pp.~348--452, March  1984.

\bibitem{ZL78}
J.~Ziv and A.~Lempel, ``Compression of individual sequences via 
variable-rate coding,''
{\em IEEE Trans.~Inform.~Theory\/},
vol.~IT--24, no.~5, pp.~530--536, September 1978.
\end{thebibliography}
\end{document}